\documentstyle[12pt,aasms4,epsf]{article}

\lefthead{ HAYASHI, SHIBATA \& MATSUMOTO}
\righthead{TWIST INJECTED MODEL OF X-RAY FLARES AND OUTFLOWS}

\begin{document}

\title{X-RAY FLARES AND MASS OUTFLOWS DRIVEN BY MAGNETIC INTERACTION
BETWEEN A PROTOSTAR AND ITS SURROUNDING DISK}

\author{M.~R. HAYASHI\altaffilmark{1},
K. SHIBATA\altaffilmark{2},}

\and

\author{R. MATSUMOTO\altaffilmark{3,4}}

\altaffiltext{1}{Department of Physics, Graduate School of Science and
Technology, Chiba University,  Inage-ku, Chiba 263, Japan ;
hayashi@spica.c.chiba-u.ac.jp}  
\altaffiltext{2}{National Astronomical Observatory, Mitaka, Tokyo 181,
Japan} 
\altaffiltext{3}{Department of Physics, Faculty of Science, Chiba University,
Inage-ku, Chiba 263, Japan}
\altaffiltext{4}{Advanced Science Research Center, JAERI, Naka, Japan}

\begin{abstract}
We propose a model of hard X-ray flares in protostars observed by
ASCA satellite. Assuming
that the dipole magnetic field of the protostar threads the
protostellar disk, we carried out 2.5-dimensional magnetohydrodynamic
(MHD) simulations of the disk-star interaction. The closed magnetic
loops connecting the central star and the disk are twisted by the
rotation of the disk. As the twist accumulates, magnetic loops expand
and finally approach to the open field configuration. A current sheet
is formed inside the expanding loops. In the presence of resistivity,
magnetic reconnection takes place in the current sheet. Outgoing
magnetic island and post flare loops are formed as a result of the
reconnection. The time scale of this `flare' is the order of the
rotation period of the disk.  The released magnetic energy partly goes
into the thermal energy and heats up the flaring plasma up to
$10^8$ K. The length of the flaring loop is several times of the
radius of the central star, consistent with observations. 
The speed of the hot plasmoid ejected by the reconnection is $200-400
$ km s$^{-1}$ when the footpoint of the loop is at 0.03 AU from 1
M$_\odot$ protostar. The hot plasma outflow can explain the
speed and mass flow rate of optical jets. Dense, cold,
magnetically accelerated wind 
($v \sim 150-250$ km s$^{-1}$) 
emanates from the surface of the disk along the partially open
magnetic field lines threading the disk. This dense, cold wind may
correspond to high velocity neutral winds. 

\end{abstract}


\section{INTRODUCTION}

Observations by ASCA revealed that protostars are strong hard X-ray 
emitting sources. By using ASCA, Koyama et al. (1994,1996a,b) carried
out a systematic survey of hard X-ray sources in molecular
clouds. They detected hard X-rays (4-10keV) associated with  Class I
IR sources (protostars). A protostar (R1 in Coronet cluster) exhibited
a powerful X-ray flare whose spectrum showed significantly higher
temperature ($T \sim 10^8$ K) than X-ray flares in T Tauri stars which is
usually $\sim (1-2) \times 10^7$ K (Koyama et al. 1996b). The X-ray
luminosity in the 2-10keV band is $\sim 10^{31}$ erg s$^{-1}$ and the
e-folding time of the flare is $\sim 5 \times 10^4$ sec (Koyama et
al. 1996b). Three X-ray emitting class I sources in Rho Oph cloud also
showed significantly hard X-rays. One of them showed flare like time
variabilities (Koyama et al. 1994).  

Although the analogy with solar flares may be appropriate for such
X-ray flares in protostars, they differ from solar flares in the
following points: (1) The total energy released by the protostellar
flare ($10^{35}-10^{36} $ erg) is $10^5$ times larger than solar
flares. (2) The spectrum of the X-ray emission in protostellar flares
is harder than solar flares, indicating the existence of 8 keV
plasma, and (3) The size of the X-ray emitting region is the order of
$10^{11}$ cm (Koyama 1995 private communication). If the emission comes
from magnetic loops, their size should be several times larger than
the protostar. 

It has long been recognized that solar flares are often associated
with sheared magnetic fields. 
Miki\'c \& Linker (1994) and Linker \&  Miki\'c (1995) carried out MHD
simulations of  
the dynamical evolution of magnetic arcades in spherical geometry under 
shearing motions imposed on the line-tied footpoints of an initially current 
free arcade. When the amount of the footpoint shear is sufficiently large,
the magnetic arcades begin to inflate. In the presence of finite resistivity,
magnetic nonequilibrium leads to the magnetic reconnection and plasmoid 
ejection in the current sheet developed inside the expanding arcades.

In the innermost region of protostellar disks, the magnetic fields of
the protostar will affect the dynamical evolution of the disk (e.g.,
Camenzind 1990). 
Since the rotation of the disk continuously injects
helicity into the closed magnetic fields connecting the central star
and the disk, we expect magnetic nonequilibrium and resulting
magnetic reconnection similar to those in solar flares (see also Ghosh
\& Lamb 1979, 1991; Zylstra 1988; Lynden-Bell \& Boily 1994). 
Recently, the 
magnetic interaction between a central star and its 
surrounding disk has attracted a renewed interest in relation to the
activity of young stars (Shu et al. 1994, Hirose 1994, Lovelace et
al. 1995). In order to explain why T Tauri stars have rotation  
rates well below the break-up despite the high accretion rate,
Shu et al. (1994) developed a steady ``X-wind model'' which emanates from the
point where the disk is corotating with its central star. Through the
magnetic interaction between the central star and the inner edge of
the disk where they assumed to be the corotation point, excess angular
momentum brought in through the disk are extracted by X-winds.   
Subsequently, Ostriker \& Shu (1995) analyzed a steady state
configuration of the magnetosphere between the central star and the
disk and discussed how dipole fields become open and generate the 
``helmet streamer'' by magnetic reconnection.

In the steady X-wind model, Shu et al. (1994) assumed that the
dipole magnetic fields of the central star do not thread the disk
except the inner edge where they assumed an enhanced magnetic
diffusivity. It is possible, however, that the magnetic fields of the
central star thread the disk either by enhanced turbulent magnetic
diffusivity inside the disk or by the recoupling of the magnetic field
and the disk matter due to the enhanced ionization rate near the
central star.    

In this Letter, we present the results of nonlinear, time-dependent MHD
simulations of the interaction between the dipole magnetic field of
the central star and the rotating disk threaded by the dipole
field. 
\section{NUMERICAL MODELS}

We solve the resistive MHD equations in cylindrical coordinate
$(r,{\phi},z)$ by applying a modified Lax-Wendroff scheme 
with artificial viscosity (see Shibata 1983
and references therein). Axial symmetry is assumed. 
The effects of viscosity, radiative cooling  and rotation of the
central star are neglected. We take the z-axis parallel to the
rotation axis of the disk and assume that the magnetic dipole moment
is aligned with the rotational axis.

Exact solutions of the disk in rotational equilibrium are obtained 
under the simplifying assumptions for the distributions of angular momentum
$L = L_0 r^a$ and pressure $P=K \rho^{1+1/n}$ where we adopted $a=0.494$
and $n=3$. The equation of force balance 
is integrated to give
\begin{equation}
-{GM\over {(r^2 + z^2)^{1/2}}} + {1\over{2(1-a)}}
L_0^2 r^{2a-2} + (n+1){P \over {\rho}} = -{GM \over {r_0}}
+{1 \over {2(1-a)}} L_0^2 r_0^{2a-2} + (n+1) {{P_0} \over {\rho_0}},
\end{equation}
where $M$ is the mass of the central star and 
the subscript 0 denotes quantities at the pressure maximum 
$(r_0,0,0)$.
The mass distribution inside the disk
is determined by specifying the parameters $a$, $n$ and $P_0/\rho_0$.
We use the normalizations
$r_0=v_{K0}=\rho_0=1$ where $v_{K0}=(GM/r_0)^{1/2}$ is the Keplerian 
rotation speed at $r=r_0$. The unit of time is $\tau_0$=$r_0/v_{K0}$.
The mass distribution outside the disk is assumed to be that of the
spherical isothermal corona surrounding the central star.
The corona is assumed to be static at $t=0$.
The coronal density at $(0,0,r_0)$ is $\rho_h=2 \times
10^{-5}\rho_0$. The coronal temperature $T_h$ is $T_h$=1.667 $\times
10^{2}T_0$, where $T_0$ is the disk temperature at pressure maximum. 
The total simulation region in the $r-z$ plane is taken to be
$0 \le r \le 13.1r_0$ and $0 \le z\le 11.0r_0$. The number of the mesh
points used in our calculation is  $(N_{r},N_{z})=(305,364)$. The grid 
spacings increase with $r$ and $z$. 

We use free boundary conditions at the outer boundaries. At the
equatorial plane ($z=0$), we assume a symmetric boundary condition.
We introduce a thin shell-like damping region (stellar surface)
between $r_{in}=(r^2+z^2)^{1/2}=0.336r_0$ and $r_s=0.528r_0$ where all
quantities except magnetic fields are damped to the initial values.
Alfv\'en waves can be reflected in this layer. All quantities are set
to the initial values in stellar interior ($r \le r_{in}$).
Meanwhile magnetic fields are stress free at the outer
boundaries. 

As for resistivity, we assume anomalous resistivity 
$\eta(r,z)=\eta_0(v_d/v_c-1.0)^2$ when $v_d = |{\bf J}|/\rho$ exceeds the
critical value $v_c$ where {\bf J} is the current density.  When $v_d <
v_c$, $\eta(r,z)$ is set zero. 
This type of anomalous resistivity
has been used in simulations of magnetic reconnection between the 
emerging flux and the overlying magnetic fields in the solar corona
(see e.g., Shibata, Nozawa \& Matsumoto 1992, Yokoyama \& Shibata 1995). 

\section{NUMERICAL RESULTS}

The results of the simulation of a typical model are shown in Figure 1
and Figure 2.
The model parameters are 
$A_{1}=v_{s0}^2/(\gamma v_{K0}^2)=2.0 \times
10^{-3}$, $A_{2}=v_{A0}^2/v_{K0}^2=2.0 \times 10^{-4}$ where $v_{s0}$
and $v_{A0}$ are the sound speed and the Alfv\'en speed at
$(r_0,0,0)$, respectively. 
The adiabatic index $\gamma$ is 5/3.
We use the anomalous resistivity with $v_c=1000$ and
$\eta_0=0.01$ in non-dimensional units. The plasma $\beta$ in the
disk at the pressure maximum is $\beta_0=8\pi P_0/B_0^2=2A_1/A_2=20$.

Solid curves in Figure 1 show 
the magnetic field lines projected onto the $r-z$ plane. The 
color scale
shows the temperature. The arrows show velocity vectors in the $r-z$ plane. 
The magnetic 
field lines connecting the central star and the disk are twisted by the 
rotation of the disk. As the magnetic twist accumulates, the magnetic loops 
begin to inflate. 
They expand quasi-statically in the
early stage but later ($t> 2 \tau_0$), they expand dynamically.
A current sheet is formed inside the expanding loops and the current
density exceeds the threshold for the 
onset of anomalous resistivity. Around $t=2.4\tau_0$, magnetic
 reconnection takes place in the current sheet. Outgoing magnetic
island and post flare loops are formed after the reconnection 
($t=2.68\tau_0$). Hot plasmas created by the Joule heating and
compressional heating are confined in the outgoing magnetic
island. The speed of outflowing hot plasma is the order of the local
Alfv\'en speed $v_A\sim (2-5) v_{K0}$. 
Crest-shaped fast MHD shock waves are formed in front of the expanding
magnetic loops. 
By parametric survey, we confirmed that when the coronal density is lower
the speed of reconnection flow and temperature are higher because
local Alfv\'en speed increases.  

Inside the magnetic island, magnetic field lines are almost
toroidal. Since they are detached from the star and the disk, the
mass and the total angular momentum of the rotating toroid is
conserved. By equating the 
centrifugal force $l^2/r^3$ where $l$ is the specific angular momentum
with the radial component of gravity $GMr/(r^2+z^2)^{3/2}$, we get $r/z
\rightarrow 0$ as $z \rightarrow \infty$ ; the plasmoid is collimated
toward the rotational axis. The azimuthal magnetic
fields also helps collimating the outflow by
self-pinching effect (see Uchida \& Shibata 1985; Shibata \& Uchida
1986). Although the computational domain in our simulation is too
small to show the collimation, the velocity vectors at $t=4.0\tau_0$
show a tendency to point toward the polar directions.  

The magnetic field lines near the disk have small angle with respect 
to the equatorial plane. This configuration is favorable for the 
appearance of magneto-centrifugal winds from accretion disks
(e.g., Blandford \& Payne 1982). 
The speed of the disk wind is the order of the Keplerian rotation 
speed at the footpoint of the field lines on the disk. Since the
density is high, the mass  
outflow rate of the cold disk wind is several times larger 
than that of the hot plasma jet driven by the  reconnection.

Figure 2 shows the density distribution (gray scale) and isocontours of
magnetic energy density. Just after the reconnection, the density of the
reconnection region decreases due to the heating and expansion. During
the later evolution, the density increases to $\rho \sim$10$^{-5} 
\rho_0$, because the colder gas that flows out of the disk refills the
region. This justifies rather high initial coronal density ($\rho_h \sim$
2 $\times$ 10$^{-5} \rho_0$) in our model.

Figure 3 shows the time history of various energies.
The solid curve in the lower panel shows the total magnetic energy
$E_{mag}$ in the simulation region. The dotted curve shows the
magnetic energy $E_{h,mag}$ of hot plasma ($T>1.2T_h$). The magnetic
energy increases almost linearly with time until magnetic reconnection
takes place. The kink around $t=2.4\tau_0$ is due to the magnetic
reconnection. The magnetic energy of hot component saturates and
slightly decreases. 
In the MHD simulations of solar coronal magnetic loops which subject
to the helicity injection, Kusano, Suzuki \& Nishikawa (1995) noted
similar kink and explained it by extending the Woltjer \& 
Taylor's minimum energy principle (Woltjer 1958, Taylor 1974).  

The solid curve and the dotted curve in the top panel of Figure 
 3 show thermal energy $E_{h,th}$, and the
kinetic energy $E_{h,kin}$ of hot plasma ($T > 1.2 T_h)$,
respectively. Rotational energy is included in $E_{h,kin}$. The
kinetic energy slowly increases during quasi-hydrostatic evolution ($t
< 2\tau_0$). The kinetic 
energy increases more steeply when the equilibrium is lost
($t \sim 2.0\tau_0$). Since the magnetic loops expand rapidly,
the rate of increase in the coronal thermal energy $E_{h,th}$
decreases during $t=2.0-2.4\tau_0$. The coronal thermal energy 
increases more steeply after magnetic reconnection at 
$t=2.4\tau_0$. These curves clearly show the bifurcation of the
system into a different energy state after magnetic reconnection.

It should be noted that the energy injected into the corona through the
rotation of the disk 
$(E_{inj}=E_{mag}+E_{h,th}+E_{h,kin})$ increases
almost linearly with time even after magnetic reconnection because
magnetic helicity is continuously injected from the disk. 
At $t=3.0\tau_0$, about 70\% of the injected energy
goes into the magnetic energy, 18\% to the thermal energy, and
12 \% to the kinetic energy. The thermal energy of hot plasma at
$t=3.0\tau_0$ is about 40\% of the initial total magnetic energy
(15\% of the total magnetic energy at $t=3\tau_0$). 

\section{DISCUSSION}

In this Letter, we demonstrated that if the dipole magnetic field of
the central star is connected to the rotating disk which is not
co-rotating with the central star, the dipole magnetic fields can
disrupt by imposed twists within one rotation of the
disk. We schematically summarize our numerical results in Figure
4. Our 2.5D nonlinear MHD simulation confirmed the previous 
expectation from the equilibrium calculations by Zylstra (1988) that
magnetic reconnection will be triggered by the loss of equilibrium due
to the helicity injection. We also showed that hot plasmoids are ejected
in bipolar directions with velocity 2$-$5 times the
Keplerian rotation speed around the inner edge of the disk. Plasma
heating occurs both by the Joule heating in the current sheet
developed inside the inflating magnetic loops and by shock waves
created by magnetic reconnection. The plasma temperature can rise up
to the temperature determined by $v_s \sim v_A \sim 2-5v_K$. 
Since $v_K \sim 170$ km s$^{-1}$ at $r=0.03$ AU around 1 $M_{\odot}
$ star, the temperature of the plasmoid and the post flare loop can be
$10^7 - 10^8$ K. 
According to the observation of an X-ray flare from a protostar
(Koyama et al. 1996b), the emission spectrum of protostellar flare
extends up to 10keV. This result is consistent with the above  plasma
temperature expected from magnetospheric flares in protostars. 

Observation of X-ray flares by ASCA (Koyama et al. 1996b) showed that
the e-folding time of the flare is $\sim 5 \times 10^4$ sec and the total
energy of the flare is $10^{36}$ erg. The size of X-ray emitting
region derived from the emission measure ($\sim 10^{54}$ cm$^{-3}$) and
flare decay time is the order of $10^{11}$ cm. 
The number density of the flaring plasma is $n_{flare} \sim 10^{10.5}$
cm$^{-3}$. This rather high density can be explained by the dynamical
refillment of the corona by the mass outflow from the disk (see
Figure 2). When the X-ray emitting plasmas are confined in loops,
their size should be larger than the radius of the protostar. Our
numerical results are consistent with this observational constraint
because hot plasmas are created in gigantic loops several times larger
than the stellar radius. The mass flux of the hot plasma outflow
obtained from simulation is 
\begin{eqnarray}
{\dot M} \sim 10^{-8} \, \biggl({r_0 \over 0.03  \, {\rm AU}} \biggl)^2
\biggl({v_{K0}\over 170 \, {\rm km} \, {\rm s}^{-1}} \biggl) 
\biggl({n_{flare} \over 3 \cdot 10^{10} \, {\rm cm}^{-3}} \biggl)
\,  M_{\odot} \, {\rm yr}^{-1}.
\nonumber
\end{eqnarray}
\noindent
After the X-ray emitting gas is cooled down (cooling time $\sim$ 1
day), it will be observed as 
optical jets. The mass flux of the hot plasma is consistent with the
observation of optical jets. 

Next, let us discuss whether the magnetospheric flares can release 
the observed total energy of X-ray flares. In our simulation, initial 
magnetic energy is thermalized within one rotation  
period ($\sim 10^5$ sec at 0.03 AU). If the size of the flaring region
is $d \sim 10^{11}$ cm and the mean strength of magnetic field in the
flaring region is $B \sim 100-200$ G, the released energy $L\sim ({B^2
/{8 \pi}}) \, d^3 \sim 10^{36}$ erg is the order of the observed luminosity.

We checked the dependence of numerical results on resistivity. When
the threshold for anomalous resistivity $v_c$ is larger, the plasma
temperature and the speed of reconnection flow are larger because
magnetic reconnection does not take place until large amount of
magnetic energy is accumulated. Since the speed of the jet is larger,
the magnetic island is deformed into a ``heart shape'' as shown in
Figure 1. This feature has also been noted by Choe (1995) when he
simulated the solar magnetic arcades under photospheric shearing
motion. When the resistivity is uniform or $v_c$ is small, the
magnetic island is more circular. 

We would like to comment on the relation between our model and the X-wind model
by Shu et al. (1994). 
They considered that the magnetic fields 
of the central star can thread the disk only in the prescribed small diffusive 
transition zone near the point where the disk corotates with the
central star.  
Interior to this radius, magnetic field lines are closed and connected to 
the surface of the star. Along the open magnetic field lines, X-winds
emanate from the small diffusive zone around the inner edge of the
disk. This configuration is similar 
to the final stage of our simulation in which the magnetic field lines 
after reconnection become partially open, and dense, cold winds flow
out from the 
disk along the partially open field lines. Although we cannot continue
the simulation until steady state is reached, we could show some essential
components of the X-wind model. Of course, we need to include the effect
of the stellar rotation in order to study the angular momentum exchange
between the star and the disk. We want to extend our study in such direction
in near future.

\vspace*{0.1cm}
\acknowledgments
\noindent
We thank Prof. K. Koyama for providing us the results of their hard
X-ray observation by ASCA prior to publication. We would like to thank
Dr. S. Miyaji, Dr. Masahiko Hayashi, Prof. T. Tajima, Dr. J.C.L. Wang,
Dr. T. Kudoh and Dr. S. Hirose for their helpful discussions.
Prof. F. Shu is  
acknowledged for introducing us the X-wind model during his visit to
the National Astronomical Observatory, Japan. We also thank
Prof. F. Lamb for pointing out the related work by Zylstra (1988).
Numerical computations were carried out by using Fujitsu VPP500 at the
Institute of Space and Astronautical Science and VPP500 at the Japan
Atomic Energy Research Institute. This work is supported in part by
the grant of Ministry of Education and Culture, Japan (07640348).

\clearpage



\begin{figure}
\caption{The result of the numerical simulation for a model
with $A_1=v_{s0}^2/(\gamma v_{K0}^2)=2.0 \times
10^{-3}$,$A_2=v_{A0}^2/v_{K0}^2 = 2.0 \times 10^{-4}$ and $\rho_h
/\rho_0=2.0 \times 10^{-5}$.
The color scale shows the temperature normalized by $T_k$=$\mu
v_{K0}^2/ {\sl R}$, here $\mu$ is mean molecular weight and {\sl R} is
gas constant. The initial coronal temperature is $T_h/T_k$=1/3.
Solid curves denote magnetic
field lines. The region of closely spaced field lines is included to
identify the reconnection region in the subsequent frames. Arrows
depict velocity vectors in $r-z$ plane. The unit velocity is shown by
arrows at the top right of each panel. 
 \label{fig1}}
\end{figure}

\begin{figure}
\caption{Time variation of the density distribution. Color scale shows the
density. Solid curves denote isocontours of magnetic energy
density. Arrows depict the velocity vector. \label{fig2}}
\end{figure}

\begin{figure}
\caption{Time history of various energies normalized by the initial
magnetic energy. Upper panal shows the thermal energy $E_{h,th}$
(solid curve) and kinetic energy $E_{h,kin}$ (dotted curve) of hot
component, 
$$E_{h,th}\!=\!2 \pi \! \int_0^{z_{max}} \! \int_{r_s}^{r_{max}} \!
[P/(\gamma - 1)] r dr \!dz,\quad
E_{h,kin}\!=\! 2 \pi \! \int_0^{z_{max}} \!
\int_{r_s}^{r_{max}} \! (\rho v^2/2)r dr \!
dz.$$
The lower panel shows the total magnetic energy $E_{mag}$ (solid curve)
and that of the hot component $E_{h,mag}$ (dotted curve). \label{fig3}}
\end{figure}

\begin{figure}
\caption{A schematic picture of numerical results. Hot plasma
jet ejected from the flaring region corresponds to the optical jet.
Cold, dense wind emanating from the disk may explain high velocity
neutral winds. \label{fig4}}
\end{figure}

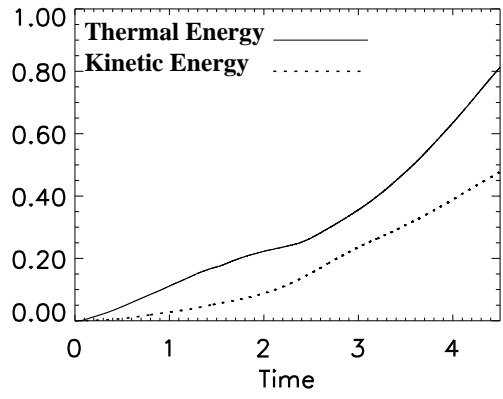
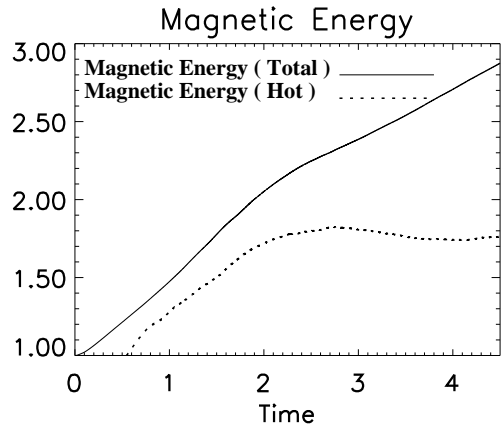
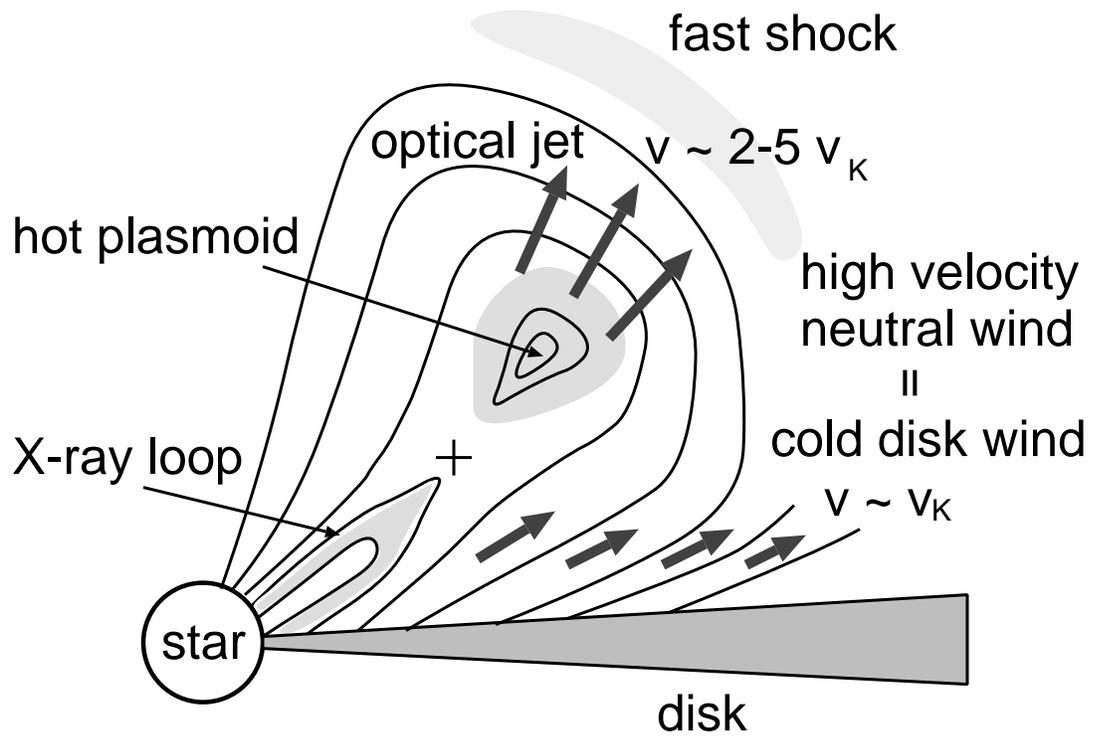
\end{document}